\documentclass[lettersize,journal]{IEEEtran}
\usepackage{amsmath,amsfonts}
\usepackage{algorithmic}
\usepackage{algorithm}
\usepackage{array}
\usepackage[caption=false,font=normalsize,labelfont=sf,textfont=sf]{subfig}
\usepackage{textcomp}
\usepackage{stfloats}
\usepackage{verbatim}
\usepackage{graphicx}
\usepackage{cite}
\usepackage{amsmath,amsfonts}
\usepackage{algorithmic}
\usepackage{algorithm}
\usepackage{array}
\usepackage[caption=false,font=normalsize,labelfont=sf,textfont=sf]{subfig}
\usepackage[normalem]{ulem}
\usepackage{multirow}
\usepackage{tabularx}
\usepackage{bm}
\usepackage{xurl}
\usepackage{hyperref}
\usepackage{array} 
\usepackage{booktabs}   
\usepackage{siunitx}    

\newcommand{\Nperc}[2]{#1}

\begin{document}

\title{Inter-Speaker Relative Cues for Two-Stage \\ Text-Guided Target Speech Extraction}

\author{Wang Dai, Archontis Politis,~\IEEEmembership {Member, IEEE}, and Tuomas Virtanen,~\IEEEmembership{Fellow, IEEE}
\thanks{Wang Dai, Archontis Politis and Tuomas Virtanen are with the Audio Research Group, Signal Processing Research Center, Tampere University, Finland. (e-mail: wang.dai@tuni.fi; archontis.politis@tuni.fi; tuomas.virtanen@tuni.fi)}
}

\markboth{Journal of \LaTeX\ Class Files,~Vol.~14, No.~8, August~2021}%
{Shell \MakeLowercase{\textit{et al.}}: A Sample Article Using IEEEtran.cls for IEEE Journals}


\maketitle

\begin{abstract}
This paper investigates the use of relative cues for text-based target speech extraction (TSE).
We first provide a theoretical justification for relative cues from the perspectives of human perception and label quantization, showing that relative cues preserve fine-grained distinctions that are often lost in absolute categorical representations for continuous-valued attributes.
Building on this analysis, we propose a two-stage TSE framework in which a speech separation model first generates candidate sources, followed by a text-guided classifier that selects the target speaker based on embedding similarity. Within this framework, we train two separate classification models to evaluate the advantages of relative cues over independent cues in case of continuous-valued attributes, considering both classification accuracy and TSE performance.
Experimental results show that (i) relative cues achieve higher overall classification accuracy and improved TSE performance compared with independent cues; (ii) the proposed two-stage framework substantially outperforms single-stage text-conditioned extraction methods on both signal-level and objective perceptual metrics; and {(iii) in controlled target-selection comparisons, the all cues, random cues, and temporal order settings outperform the strong enrollment-audio selection based on ECAPA-TDNN speaker embeddings, while language and distance cues remain competitive. Other textual cues, including gender, loudness, speaking duration, pitch, speaking rate, and transcription, all exceed 90\% selection accuracy. In the WHAM!-noise evaluation, textual cues also show strong noise robustness, while all cues, random cues, and gender are especially robust. These results demonstrate the practical value of textual cues when enrollment speech is unavailable or cannot be used due to privacy concerns.} Further analysis reveals notable differences in discriminative power across cue types, providing insights into the effectiveness of different relative cues for TSE.
\end{abstract}


\section{Introduction}
\IEEEPARstart{T}SE systems leverage auxiliary information—such as speaker embeddings derived from an enrollment speech sample, spatial location, visual cues from lip movements, and textual descriptions of the target speaker’s attributes—to enable the TSE model to selectively attend to and extract the desired voice in a mixture of multiple speakers \cite{zmolikova2023neural,11129612}. Applications such as smart assistants, hearing aids, and meeting transcription benefit from TSE’s ability to isolate the desired voice, enhancing both accuracy and usability by leveraging speaker-specific cues \cite{zmolikova2023neural, 11129612, sato24_interspeech, alcalapadilla25_interspeech}.

Recent years have witnessed rapid advances in target speech extraction, driven by progress in deep learning, self-supervised pretraining, and multimodal modeling. 
Early works in speaker-dependent extraction exploited the idea of conditioning the separation network on speaker identity, giving rise to architectures such as SpeakerBeam \cite{8736286} and VoiceFilter \cite{Wang2019}. These models employ a reference-guided framework, where a short clean utterance of the target speaker provides a speaker embedding that modulates the separation network to enhance the target voice.

Subsequent developments incorporated pretrained speaker encoders, such as D-Vectors\cite{6854363}, X-Vectors\cite{8461375}, and ECAPA-TDNN\cite{desplanques20_interspeech}, to obtain robust speaker embeddings, decoupling speaker modeling from the extraction network. 
Moreover, multi-modal approaches integrating audio and video cues have gained attention\cite{pan2020muse,li2022vcse,tao2025seanet,wu2024unified}, especially for audiovisual datasets like AVSpeech\cite{ephrat2018looking} or VoxCeleb\cite{nagrani2020voxceleb}. These methods align visual speech movements or lip information with audio stream\cite{assael2016lipnet,li2023facelipav}, improving robustness in visually grounded TSE scenarios.

Despite these advancements, several challenges remain, including robustness to unseen speakers, limited availability of clean target references in real-world scenarios, and mismatch between the enrollment speech and the mixture input regarding recording conditions, speaking style, or acoustic environment.
Beyond acoustic/visual-based guidance, recent research has explored text-guided extraction. These models use a text query (prompt) describing the speech attributes of a target speaker as a condition to guide the extraction of the desired speaker’s speech signal from a mixture, enabling extraction when no enrollment utterance of the target speaker is available.
Compared to other cues, text offers greater accessibility, enabling users to provide instructions without recording audio or video. It also provides greater flexibility, as it can convey a wide range of target speaker characteristics, including linguistic cues (e.g., speech transcription), paralinguistic cues (e.g., gender), and acoustic cues (e.g., loudness of speech). In addition, text offers stronger privacy by avoiding the transmission or storage of voice data or other media.
 
Hao et al. were the first to successfully introduce text-based TSE that used a LLM for extracting semantic information from text queries and integrating it into a speech separation network \cite{hao2025typing}. Subsequently, Huo et al. proposed a flexible text-guided TSE framework and introduced the TextrolMix dataset, which provides rich textual descriptions of speech attributes, including speaker identity, emotion, pitch, gender, accent, and speaking rate \cite{huo2025beyond}. Similarly, Seki et al. developed a TSE model conditioned on text prompts describing paralinguistic and non-linguistic attributes such as gender, pitch, speaking speed, and loudness, demonstrating that it can overcome the limitations of existing audio-based TSE methods \cite{seki2025language}. { Closely related to transcription-based textual cues, Li et al. proposed DAE-TSE, a keyword-guided TSE framework that detects partial transcriptions, attends to the corresponding speaker, and extracts the target speech \cite{li2026detect}. Their work specifies the target using words uttered by the speaker, whereas the present work focuses on relative perceptual and acoustic attributes that distinguish speakers in a mixture.}

Previous text-based TSE studies typically represent continuous-valued speech attributes, such as speaking rate, using a small number of discrete categories (e.g., slow, normal, fast) without considering inter-speaker relations. This is usually done by first estimating each speaker's attribute values in a mixture and then quantizing them into predefined categories. Such coarse quantization can obscure meaningful inter-speaker differences, as within-category variation is discarded, reducing discriminative information. Consequently, speaker-independent cues lose finer-grained distinguishing information when the target and interfering speakers have subtle attribute differences but are assigned to the same category.

To overcome this limitation, our prior study introduced inter-speaker relative cues, which explicitly describe differences between the target and interfering speakers \cite{dai25b_interspeech}. By explicitly encoding inter-speaker comparisons, such as “Speaker A has a faster speaking rate than Speaker B”, relative attribute cues preserve these contrasts, resulting in more discriminative and perceptually aligned conditioning for target speaker extraction. The present journal manuscript is an extended version of that conference paper. In the prior work, we introduced the relative-cue formulation, constructed a two-speaker mixture dataset spanning pitch, pitch range, speaking rate, speaking duration, temporal order, age, loudness, speaker-to-microphone distance, language, gender, transcription, and emotion, and evaluated a single-stage text-guided TSE system. In this journal version, we retain the conference material as essential background and substantially extend it through theoretical justification, a novel two-stage text-guided TSE modeling paradigm, and extensive experimental validation and analysis.
Specifically, this work makes four significant contributions beyond our prior work \cite{dai25b_interspeech}.
\begin{itemize}

\item \textbf{Theoretical Justification of Relative Cues}: We provide a theoretical justification for inter-speaker relative cues by grounding them in the speech perception literature and deriving mathematical formulations from the perspectives of human perception and label quantization. This analysis establishes a principled theoretical foundation for the proposed relative-cue representation and directly motivates the subsequent empirical comparison between relative cues and speaker-independent cues, which we refer to hereafter as independent cues throughout the remainder of the paper.

\item \textbf{Proposed Two-Stage Text-Guided TSE Framework}: We propose a two-stage text-guided modeling framework that is fundamentally different from the previous single-stage text-guided approach. The two frameworks differ in both modeling methodology and optimization behavior. This design directly mitigates speaker confusion inherent in single-stage text-guided TSE systems, makes cue-wise target-selection behavior interpretable through classification accuracy, and provides a useful way to analyze the discriminative power of different textual cues.

\item \textbf{Extensive Experimental Analysis}: Based on this two-stage framework, we conduct extensive experimental validation to support the proposed formulation and analyze when and why relative cues are effective.

\item \textbf{Empirical Evaluation of the Advantages of Relative Cues}: Based on the proposed two-stage framework, we evaluate the impact of relative versus independent cues on overall target-speaker classification and the corresponding TSE performance for comparable continuous-valued attributes. This analysis provides an empirical perspective on the advantages of relative cues over independent cues.

\end{itemize}
The remainder of this paper is organized as follows. Section II introduces the inter-speaker relative cues and presents their theoretical justification. Section III describes the proposed two-stage method. Section~IV presents the dataset, training configuration, evaluation setup, experimental results, and corresponding analysis. Section V discusses limitations and future directions. Finally, Section VI concludes the paper.

\section{Inter-Speaker Relative Cues}
\begin{figure*}[!t]  
    \centering
    \subfloat[{\textrm{\fontsize{9}{10}\selectfont Independent Cue}}]{%
        \includegraphics[width=0.32\textwidth]{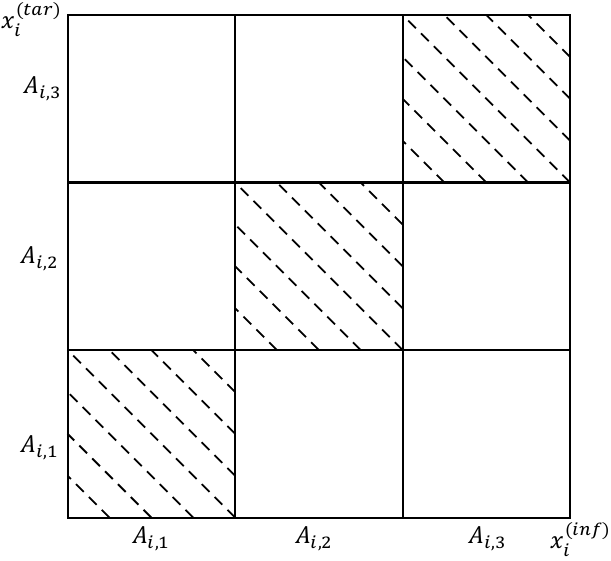}%
        \label{fig:fig1}%
    }
    \hfill
    \subfloat[{\textrm{\fontsize{9}{10}\selectfont Relative Cue for Direct Difference}}]{%
        \includegraphics[width=0.32\textwidth]{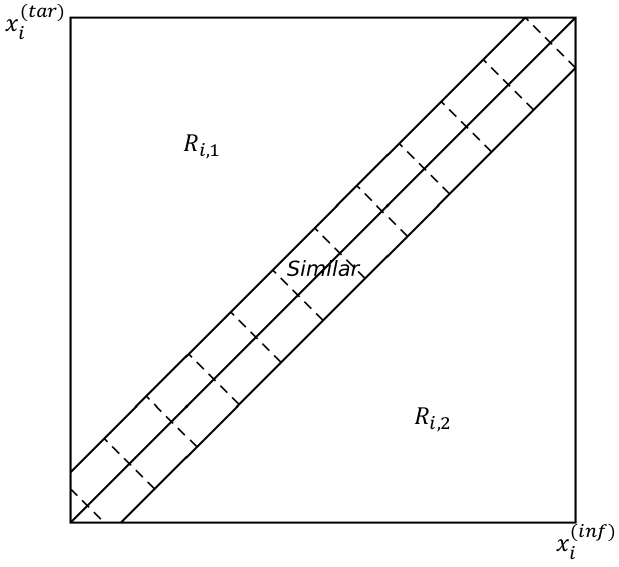}%
        \label{fig:fig2}%
    }
    \hfill
    \subfloat[{\textrm{\fontsize{9}{10}\selectfont Relative Cue for Percentage Difference}}]{%
        \includegraphics[width=0.32\textwidth]{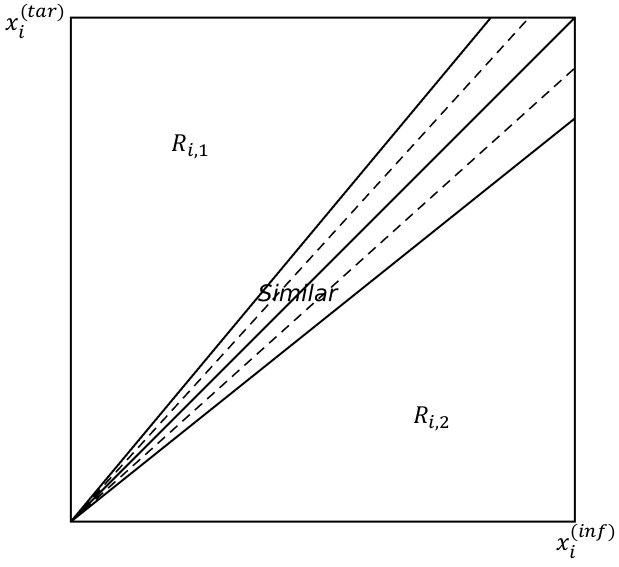}%
        \label{fig:fig3}%
    }
    \caption{Illustration of quantization schemes for continuous-valued attributes.
(a) Independent cues: the attribute value space of the target and interfering speakers, denoted by $x_i^{(\mathit{tar})}$ (y-axis) and $x_i^{(\mathit{inf})}$ (x-axis), is partitioned into a small number of coarse categories (three shown as in the example). Shaded regions indicate attribute values of the two speakers that are quantized into the same category, in which case the two speakers cannot be distinguished based on the attribute.
(b) Relative cues for direct differences: attribute values of the target and interfering speakers are compared with direct difference. The shaded \emph{Similar} region corresponds to $|\Delta x_i^{(*)}| \le \theta_i$, indicating differences below the discrimination threshold. Regions outside the shaded area represent perceptually distinct differences.
(c) Relative cues for percentage differences: attribute values of the target and interfering speakers are compared with percentage difference. The shaded \emph{Similar} region corresponds to $|\Delta x_i^{(*)}| \le \theta_i$, indicating differences below the discrimination threshold. Because the comparison is based on percentage differences, the \emph{Similar} region gradually expands from the zero point. Regions outside the shaded area represent perceptually distinct relative differences.}
    \label{fig:cue_illustration}
\end{figure*}
The proposed text-guided TSE system takes a two-speaker mixture as input along with a text prompt which consists of cues indicating the target speaker, aiming to extract the target speech signal. In this work, we employ \textbf{inter-speaker relative cues}, which describe the relative differences in speech attributes between the target speaker and the interfering speaker in the mixture. According to the value type of speech attributes, they can be categorized into \emph{discrete-valued} and \emph{continuous-valued} types.

Continuous-valued speech attributes correspond to estimations of specific speech or acoustic properties.
In this study, these attributes include:
\begin{enumerate}
    \item \textbf{mean $\bm{F_0}$}, computed as the average fundamental frequency over all voiced frames in an utterance;
    \item \textbf{$\bm{F_0}$ span}, defined as the difference between the maximum and minimum ${F_0}$ within an utterance;
    \item \textbf{speaker age}, treated as a continuous variable in years;
    \item \textbf{speaking duration}, defined as the total active speech time (in seconds) of a speaker within the mixture;
    \item \textbf{speaking rate}, measured as the number of syllables per minute in the speech signal;
    \item \textbf{RMS energy (dB)}, computed as $20 \log_{10}$ of the root-mean-square amplitude of the speech signal;
    \item \textbf{speaker-to-microphone distance}, expressed in meters; and
    \item \textbf{appearance time}, defined as the onset time of the first occurrence of a speaker’s speech in the mixture.
\end{enumerate}

Discrete-valued speech attributes correspond to categorical properties that take values from a finite set. In this study, these attributes include:
\begin{enumerate}
\item \textbf{language}, indicating the spoken language of the utterance;
\item \textbf{gender}, representing the speaker’s perceived acoustical gender category;
\item \textbf{emotional state}, describing the affective category expressed in the speech signal; and
\item \textbf{transcription}, corresponding to the linguistic content of the utterance in textual form.
\end{enumerate}

We derive each inter-speaker relative cue by comparing a single speech attribute between the target speaker and the interfering speaker in a two-speaker mixture. Each relative cue encodes the difference along one specific attribute only.
Independent cues, in contrast to relative cues, are derived solely from the attributes of an individual speaker—either the target or the interfering speaker—without accounting for inter-speaker relationships.
\subsection{Independent Cues}
The key difference between relative and independent cues lies in how continuous-valued attributes are represented.
Let $x_i^{(\mathrm{tar})}$ and $x_i^{(\mathrm{inf})}$ denote the value of a continuous-valued speech attribute $i$ (e.g., speaking rate) for the target and interfering speakers, respectively. 
\emph{Independent cues} represent each continuous-valued attribute using a small number of discrete categories. The attribute space is segmented into these categories based on the distribution of attribute values across all individual target and interference speakers in the training set.
Each attribute value $x_i$ is quantized as

\begin{equation}
\tilde{x}_i = \mathcal{Q}_i(x_i), \quad 
\tilde{x}_i \in \{\mathit{A}_{i,1}, \mathit{A}_{i,2}, \dots, \mathit{A}_{i,k_i}\},
\end{equation}

where $\mathcal{Q}_i(\cdot)$ denotes the quantization operator for the $i$-th attribute, $k_i$ is the total number of discrete categories for that attribute, and $\mathit{A}_{i,1}, \dots, \mathit{A}_{i,k_i}$ denote the corresponding categories (e.g., low, normal, and high for mean $F_0$).

The independent cue category for attribute $i$ is given by $\tilde{x}_i$. The mapping from continuous-valued attribute values $x_i$ to discrete independent cue categories $\tilde{x}_i$ is illustrated in Fig.~1(a). Due to the coarse nature of the categories, perceptually noticeable differences in the original continuous-valued attribute may be mapped to the same discrete category (as illustrated by the shaded region in Fig.~1(a)), in which case the attributes of two speakers cannot be distinguished from each other.

\subsection{Relative Cues}
In contrast, \emph{relative cues} categorize continuous-valued attributes into groups (e.g., slower, similar, and faster for speaking rate) by comparing the attribute values of the two speakers. 
For the attributes RMS energy (dB), speaker-to-microphone distance, speaker age, and appearance time in the mixture, the difference is expressed as the \emph{direct difference}

\begin{equation}
\Delta x_i^{(.)} = x_i^{(\mathrm{tar})} - x_i^{(\mathrm{inf})}.
\end{equation}

These attributes correspond to interpretable physical or temporal quantities whose perceptual relevance is largely determined by differences in their values. In human perception, changes in stimulus magnitude can be detected through difference thresholds, allowing listeners to perceive relative differences in such attributes between competing sources. Consequently, variations in these quantities can be directly perceived and interpreted without requiring an explicit baseline reference \cite{moore2012psychology,bregman1990auditory}.
For attributes such as mean $F_0$, $F_0$ span, speaking rate, and speaking duration, we employ the \emph{percentage difference}

\begin{equation}
\Delta x_i^{\%} =
\frac{x_i^{(\mathrm{tar})} - x_i^{(\mathrm{inf})}}
{\min\!\left(x_i^{(\mathrm{tar})}, x_i^{(\mathrm{inf})}\right)}
\times 100\%.
\end{equation}

These attributes represent global prosodic properties, for which human listeners are more sensitive to proportional than absolute changes \cite{Fant1970,LadefogedJohnson2014,Stevens1957}, making this formulation better aligned with human auditory perception across different baselines.

For notational convenience, we introduce the variable
\(\Delta x_i^{(*)} \in \{\Delta x_i^{(.)}, \Delta x_i^{\%}\}\)
to denote the attribute-specific difference measure.
This measure is then compared against the predefined threshold \(\theta_i\) of each attribute, which are shown in Table~I.
\begin{table*}[!t]
\caption{Continuous-Valued Speech Attributes with Our Defined Thresholds for Relative Cues and Motivating Speech Perceptual Studies}
\centering
\begin{tabularx}{\textwidth}{l c X}
\hline
\textbf{Speech Attribute} & \textbf{Threshold (\(\theta_i\))} & \textbf{Findings from Motivating Speech Perception Literature} \\
\hline

RMS energy (dB) & 3 dB &
Perception of relative loudness differences in speech is mediated by changes in loudness-related acoustic features, such as excitation strength derived from the glottal source \cite{seshadri2009perceived}. 
McShefferty et al.~\cite{mcshefferty2015just} report that the average just-noticeable difference (JND) in speech-to-noise ratio is approximately 3~dB. 
\\
\hline

Speaker-to-microphone distance & 0.5 m &
Human perception of speaker-to-microphone distance relies on relative variations in distance-dependent acoustic cues, particularly changes in the direct-to-reverberant energy ratio \cite{sheeline1983investigation, mershon1989effects, zahorik2005auditory}.
\\
\hline

Speaker age & 10 years &
Perception of speaker age is based on relative differences in age-related vocal characteristics, such as vocal tract resonance and speaking rate. Listeners typically judge speakers as ``younger'' or ``older'' rather than estimating absolute age \cite{linville2001vocal, skoog2015can, winkler2007influences}.
\\
\hline

Mean $F_0$ & 6\% &
Pitch perception is governed by relative changes in F0. Listeners’ sensitivity to differences in F0 changes in speech and found that for many listeners, average JNDs were around 1.5–2 semitones (6-12\%) \cite{t1981differential}.
\\
\hline

$F_0$ span & 25\% &
Perception of pitch range ($F_0$ span) depends on relative variation in pitch-related features. Wider pitch ranges are often associated with high arousal or positive affect, while narrower ranges indicate low arousal \cite{jusczyk1997discovery, banse1996acoustic, scherer2003vocal, juslin2003communication}.
\\
\hline

Speaking rate & 15\% &
Humans perceive differences in speaking rate through relative tempo changes, with a JND of approximately 5\% of the base rate \cite{quene2007just}. Descriptors such as ``much slower'' or ``faster'' reflect this relative perception \cite{feldstein1981perception, rietveld1987perceived}.
\\
\hline

Speaking duration & 15\% &
Perception of speech duration follows Weber’s law, whereby the JND scales proportionally with the original duration, indicating reliance on relative temporal differences \cite{haigh2021role}.
\\
\hline

Appearance time & 0.1 s &
\\
\hline

\end{tabularx}

\vspace{0.3em}
\noindent\parbox[t]{\textwidth}{\footnotesize
}
\end{table*}
Some thresholds are motivated by directly adopting suggested JNDs or by adjusting them to larger values based on findings from prior psychoacoustic and speech perception studies. The remaining thresholds, which either lack suggested JNDs or have no related studies, are defined by us\footnote{These thresholds are largely chosen in an ad hoc manner, as no perceptual studies have used exactly the same experimental setup.}.

The relative cue category \(r_i\) for attribute $i$ is assigned as

\begin{equation}
r_i =
\begin{cases}
\mathit{R}_{i,1}, & \Delta x_i^{(*)} > \theta_i, \\[2pt]
\mathit{Similar}, & |\Delta x_i^{(*)}| \le \theta_i, \\[2pt]
\mathit{R}_{i,2}, & \Delta x_i^{(*)} < -\theta_i,
\end{cases}
\end{equation}

where \(\mathit{R}_{i,1}\), \(\mathit{Similar}\), and \(\mathit{R}_{i,2}\) denote relational categories corresponding, for example, to longer, similar, and shorter speaking duration relative to the interfering speaker. 
This mapping is illustrated in Fig. 1(b) and Fig. 1(c).
The distribution of these relative cues under different threshold settings in the test set (based on the simulated dataset described in Section IV) is presented in Table II. The total number of samples is nearly balanced across cue types, except for speaker age, for which labeled samples are limited in our dataset.

\begin{table}[h]
    \centering
    \caption{Sample counts of relative categories for each continuous attribute in the test set based on defined thresholds}
    \label{tab:perceptual_thresholds}
    \renewcommand{\arraystretch}{1.2}
    \begin{tabular}{
    p{2.7cm}|
    >{\raggedleft\arraybackslash}p{1.2cm}|
    >{\raggedleft\arraybackslash}p{1.2cm}|
    >{\raggedleft\arraybackslash}p{1.2cm}
    }
        \hline
        \textbf{Continuous-valued Attribute} & \textbf{R1} & \textbf{Similar} & \textbf{R2} \\
        \hline
        RMS energy 
        & 2547 & 4992 & 2461 \\ 
        \hline
        Speaker-to-microphone distance 
        & 1441 & 7124 & 1435 \\ 
        \hline
        Speaker age
        & 198 & 605 & 197 \\
        \hline
        Mean $F_0$ 
        & 4394 & 1081 & 4515 \\ 
        \hline
        $F_0$ span  
        & 3799 & 2376 & 3813 \\ 
        \hline
        Speaking rate
        & 3331 & 2235 & 3465 \\ 
        \hline
        Speaking duration
        & 3881 & 2227 & 3892 \\ 
        \hline
        Appearance time
        & 3179 & 3511 & 3310 \\ 
        \hline
    \end{tabular}
\end{table}

\subsection{Discrete-Valued Attributes}
For discrete-valued attributes, independent cues specify the attribute value (category) of the target speaker (e.g., the language spoken) and do not consider whether it is shared with the interfering speaker. In contrast, relative cues indicate whether the target and interfering speakers share the same attribute value (category) or are distinct with respect to that attribute.
Let $d_i^{(\mathrm{tar})}$ and $d_i^{(\mathrm{inf})}$ denote the values of discrete-valued attribute $d_i$ for the target and interfering speakers, respectively, with $i$ indexing a particular attribute (e.g., language). Then the independent cue category $a_i$ for the target speaker is defined as \(a_i = d_i^{(\mathrm{tar})}\), while the relative cue category $r_i$ for the target speaker is defined as

\begin{equation}
r_i =
\begin{cases}
\text{Same}, & d_i^{(\mathrm{tar})} = d_i^{(\mathrm{inf})},\\[1mm]
d_i^{(\mathrm{tar})}, & d_i^{(\mathrm{tar})} \neq d_i^{(\mathrm{inf})},
\end{cases}
\end{equation}

so that when the attribute values (categories) differ, the relative cue retains the original categorical distinction of the target speaker.
\subsection{Text Prompt Generation}
Inter-speaker relative cues are formulated as natural language descriptions (i.e., text prompts) that explicitly capture the differences between the target and interfering speakers, and are used for both training and evaluation of the TSE systems. For each mixture, a target speaker is specified based on the dataset annotations (see Section~IV), and a corresponding prompt is generated by instantiating a predefined template with one or more relative cues. Specifically, one prompt is generated for each available individual cue, as well as for combinations of cues, either randomly selected or comprising all available cues (see details below). In the first experimental setting (Section~IV-C), cue categories are chosen to ensure clear inter-speaker contrast, thereby providing sufficient discriminability for target speaker extraction. In contrast, the second experimental setting (Section~IV-D) adopts a different definition of cue categories, allowing labels such as ``same'' or ``similar'' during testing. The template structure used for prompt generation follows the implementation described in the code\footnote{\url{https://github.com/daiwangsnr/Inter-Speaker-Relative-Cues-for-Text-Guided-Target-Speech-Extraction/blob/main/prompt_template.py}}, incorporating action-oriented verbs such as \textit{extract}, \textit{separate}, or \textit{isolate}, along with descriptive phrases that emphasize the distinguishing characteristics of the target speaker. For example, a cue vector such as (female, higher pitch, faster speaking rate) can be verbalized as: ``Please extract the female speaker with a higher pitch and faster speaking rate.''

Prompts are generated in three configurations for each mixture:
\begin{itemize}
    \item \textbf{Individual cues:} A separate prompt is generated for each available cue.
    \item \textbf{Random cues:} If at least three cues are available, a subset is randomly sampled for each mixture, with size $2$ to $n-1$ (where $n > 3$ is the total number of available cues), and used to generate a prompt. This configuration enables robust evaluation under partial and randomly combined cues.
    \item \textbf{All cues:} If at least two cues are available, a prompt is generated that incorporates all cues, enabling evaluation under fully combined cue conditions.
\end{itemize}

We also experimented with GPT-4o-mini~\cite{achiam2024gpt4} to automatically rephrase templates for greater linguistic diversity. { Although this introduced greater variation, the generated prompts sometimes contained less precise or non-canonical cue descriptions. In some cases, the intended cue was even replaced with a different attribute that was not included in the ground-truth labels. For example, “quieter loudness” could be rephrased as “higher pitch,” even though pitch was not a ground-truth cue for that sample. Such label-inconsistent semantic deviations create a prompt-distribution shift relative to the deterministic templates used to define the cue labels and may weaken the alignment between the text embedding and the intended acoustic attribute. Therefore, to control lexical ambiguity and isolate the effect of cue formulation, we retain template-based prompts in our experiments.} Table~\ref{tab:relative_cue_counts} (Section~IV) reports the number of mixtures associated with each relative cue in each subset. Some cue names differ from their corresponding speech attributes (e.g., loudness\footnote{In this work, ``loudness'' refers to RMS energy and does not strictly correspond to its conventional perceptual definition} vs.\ RMS energy, age vs.\ speaker age, pitch vs.\ mean $F_0$, pitch range vs.\ $F_0$ span, distance vs.\ speaker-to-microphone distance, temporal order vs.\ appearance time, and emotion vs.\ emotional state). These differences arise because some cue names are abbreviated, while others are expressed in more perceptually meaningful terms. Since not all mixtures include every cue (i.e., the target speaker is not necessarily characterized by all possible cues), the counts for individual cues do not sum to the total number of mixtures.
\begin{figure*}[htbp]
\centering 
\includegraphics[
    width=1.0\textwidth,
]{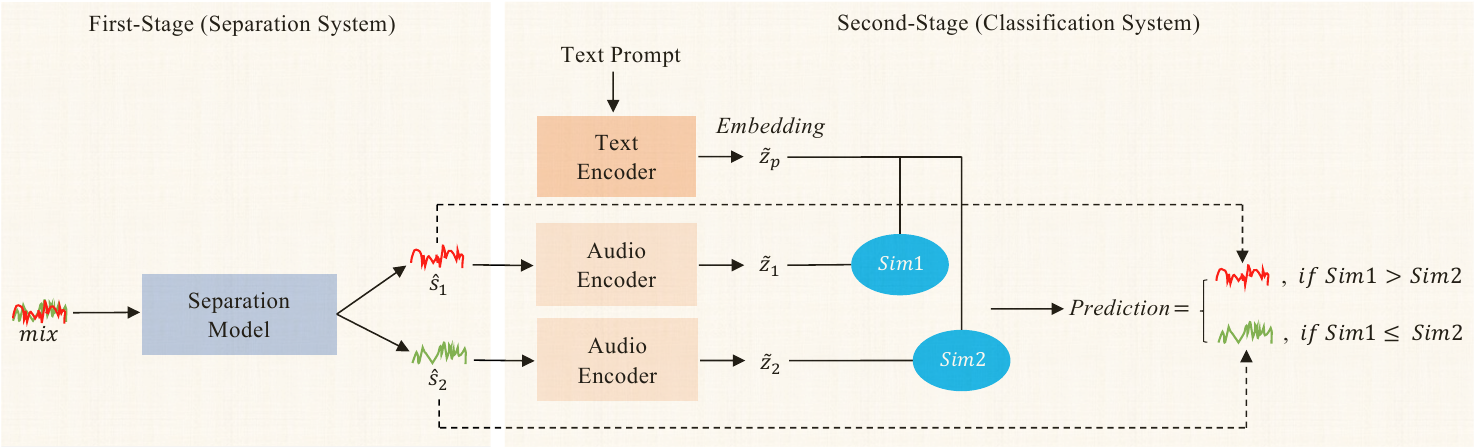}
\caption{Overview of the proposed two-stage text-guided TSE system during inference.}
\label{fig:two-stage-framework}
\end{figure*}

\section{Proposed Two-Stage Method}
In this work, we propose a two-stage approach for text-guided target speech extraction, comprising: (1) speech separation and (2) target speech classification guided by a text prompt. This design is inspired by prior two-stage target speech extraction framework, which first separate mixture speech signals that perform speech separation followed by speaker verification based on speaker embeddings derived from enrollment utterances \cite{elminshawi2024new}. Such a cascaded design \cite{elminshawi2024new} has been shown to improve TSE performance compared to single-stage TSE systems.
In this two-stage framework, the two components are trained independently, which simplifies the optimization process and allows each module to focus on a more specific objective. This modular strategy leads to a more stable and interpretable architecture than single-stage TSE approaches.
An overview of the system during inference is provided in Fig.~\ref{fig:two-stage-framework}: the input mixture is first separated into individual speech signals, after which a text-guided target selection mechanism identifies and outputs the desired speech signal.

Let the input mixture be denoted as \(mix\), which contains two speakers in the experiments of this paper. Given a text prompt \(p\) describing one or more relative cues of the target speaker, the goal is to extract the target speech signal from \(mix\). The following formulation describes the two-speaker setting used for training and evaluation.
\subsection*{Stage 1: Speech Separation}
In the first stage, the mixture is separated into individual speech signals using a speech separation model \(\mathcal{S}(\cdot)\) as

\begin{equation}
[\hat{s}_1, \hat{s}_2] = \mathcal{S}({mix}),
\end{equation}

trained with utterance-level permutation invariant training (PIT) \cite{kolbaek2017multitalker}, where \(\hat{s}_1\) and \(\hat{s}_2\) denote the estimated speech signals of the individual speakers. PIT resolves the permutation ambiguity by computing the loss for the two possible output-reference assignments and selecting the minimum. This allows the model to learn speaker-specific separation without assuming a fixed output order. The separation stage produces isolated speech signals that serve as input for the subsequent target speech classification stage.
\subsection*{Stage 2: Target Speech Classification via Text Prompt}
The second stage identifies which separated signal corresponds to the target speaker using the provided text prompt.
\paragraph{Speaker and Text Embeddings}
Each separated signal \(\hat{s}_j\) (where \(j \in \{1,2\}\)) is encoded by an audio encoder as \(\tilde{z}_j = \mathcal{A}(\hat{s}_j)\), producing a fixed-dimensional embedding.
The audio encoder consists of a pre-trained self-supervised speech model (wav2vec2-large-xlsr-53\footnote{\url{https://huggingface.co/facebook/wav2vec2-large-xlsr-53}}), followed by a pooling operation, ReLU activation, and layer normalization.

Similarly, the text prompt \(p\) is encoded into the same embedding space as \(\tilde{z}_p = \mathcal{T}(p)\), where \(\mathcal{T}\) includes an LLM-based model (LLaMA 3.2 1B Instruct~\cite{dubey2024llama3}) followed by a trainable linear projection with ReLU and layer normalization, mapping the LLM output to the speaker embedding space to align textual and speaker representations in a shared, directly comparable latent space.
\paragraph{Similarity Computation and Classification}
The similarity of the text prompt and each separated signal is measured via the cosine similarity of their embeddings as

\begin{equation}
\text{sim}_j = \frac{\tilde{z}_p \cdot \tilde{z}_j}{\|\tilde{z}_p\| \, \|\tilde{z}_j\|}, \quad j \in \{1,2\}.
\end{equation}

We define the logit as the difference between the two similarity scores, \(\text{logit} = \text{sim}_1 - \text{sim}_2\), which measures how much more similar the separated signal \(\hat{s}_1\) is to the text prompt compared to \(\hat{s}_2\). The logit is scaled by a temperature parameter \(T\) (\(T < 1\)) and mapped to a probability using the sigmoid function \(\hat{y} = \sigma(\text{logit}/T)\), where \(\hat{y} \in [0,1]\) represents the likelihood that \(\hat{s}_1\) corresponds to the target speaker described by the prompt. A temperature \(T < 1\) sharpens the output distribution, making predictions more confident (closer to 0 or 1). Empirically, we find that \(T = 0.2\) yields good performance. The predicted target speaker index is then given by

\begin{equation}
\text{pred} =
\begin{cases}
1, & \hat{y} > 0.5,\\
0, & \text{otherwise}.
\end{cases}
\end{equation}

During inference, the signal with the higher similarity score is selected as the target speech \(\hat{s}_{\text{tar}}\).
The present work focuses on two-speaker mixtures; extensions to more general source conditions are discussed as limitations and future directions in Section~V.
\paragraph{Training Label Generation and Loss}
To train the classifier, ground-truth labels are generated for each two-speaker mixture by comparing the separated signals with the reference target speech \(s_{\mathrm{ref}}\) using scale-invariant signal-to-distortion ratio (SI-SDR) \cite{le2019sdr} as

\begin{equation}
y =
\begin{cases}
1, & \text{if } \hat{s}_1 \text{ achieves higher SI-SDR with respect to } s_{\mathrm{ref}},\\
0, & \text{otherwise}.
\end{cases}
\end{equation}

These labels indicate which separated signal corresponds to the target speaker. The classifier is trained by minimizing the binary cross-entropy (BCE) loss \(\mathcal{L} = \mathrm{BCE}(\hat{y}, y)\) between the predicted probability and the generated label, encouraging the model to correctly associate the text prompt with the separated speech signal of the target speaker.

\section{Evaluation}
The evaluation aims to: 
(i) assess the advantages of the proposed two-stage framework over conventional single-stage systems and compare the effectiveness and noise robustness of different relative cues, both across cue types and against enrollment-audio-based baselines within the two-stage framework; and (ii) examine whether relative cues yield better performance than independent cues.
The following sections provide an overview of the experimental dataset and training configurations, followed by detailed descriptions of each experiment, including the setup, results, and analysis.

\subsection{Dataset}
To ensure broad linguistic and acoustic diversity, we construct audio mixtures from several public corpora across five languages, as summarized in Table~\ref{tab:source_corpora}. The corpora provide complementary annotations for language, gender, transcription, emotion, and age, enabling the construction of both discrete-valued and continuous-valued relative cues. Each mixture contains speech from two speakers--one designated as the \emph{target} and the other as the \emph{interference}--together with a text prompt describing their relative characteristics.

\begin{table}[h]
\centering

\caption{Summary of speech corpora used for mixture construction. The speaker column reports training/validation/test speakers.}
\label{tab:source_corpora}
\renewcommand{\arraystretch}{1.0}
\setlength{\tabcolsep}{4.5pt}
\begin{tabular}{p{2.22cm}|p{1.15cm}|p{1.1cm}|p{2.3cm}}
\hline
\textbf{Dataset} & \textbf{Language} & \textbf{Speakers} & \textbf{Attributes} \\ \hline
CASIA \cite{zhang2008design} & Chinese & 2/1/1 & Language, Emotion, Gender, Transcription \\ \hline
ESD \cite{zhou2022emotional} & Chinese & 6/2/2 & Language, Emotion, Gender, Transcription \\ \hline
Aishell3 \cite{shi21c_interspeech} & Chinese & 120/30/30 & Language, Gender, Transcription \\ \hline
MagicData-Read \cite{magicdata2019} & Chinese & 33/8/11 & Language, Age, Gender \\ \hline
Librispeech \cite{panayotov2015librispeech} & English & 120/30/30 & Language, Gender, Transcription \\ \hline
ESD \cite{zhou2022emotional} & English & 6/2/2 & Language, Emotion, Gender, Transcription \\ \hline
OreauFR-02 \cite{kerkeni2020french} & French & 21/7/4 & Language, Emotion, Age, Gender \\ \hline
MLS French \cite{Pratap2020MLSAL} & French & 90/18/18 & Language, Gender, Transcription \\ \hline
EmoDB \cite{burkhardt2005database} & German & 6/2/2 & Language, Emotion, Age, Gender, Transcription \\ \hline
MLS German \cite{Pratap2020MLSAL} & German & 120/30/30 & Language, Gender, Transcription \\ \hline
MLS Spanish \cite{Pratap2020MLSAL} & Spanish & 79/20/20 & Language, Gender, Transcription \\ \hline
EmoMatchSpanish \cite{garcia2024emomatchspanishdb} & Spanish & 30/10/10 & Language, Emotion, Gender, Transcription \\ \hline
\end{tabular}
\end{table}

(1) Extraction of Speech Attributes:
To provide ground-truth labels for the relative cue categories in each mixture, a set of speech attributes is extracted for every utterance. These attributes serve as the basis for constructing inter-speaker relative cues (e.g., higher pitch, faster speaking rate). All attributes are computed from the isolated clean speech signals prior to mixing, ensuring that the measurements are not affected by interference from other speakers or reverberation.

Mean $F_0$ and $F_0$ span are extracted using the Librosa pYIN library\footnote{\url{https://librosa.org/doc/main/generated/librosa.pyin.html}}.
When computing speaking duration, word- and pause-level durations are first obtained. If transcription annotations are available, these durations are extracted from time boundaries provided by the Montreal Forced Aligner\footnote{\url{https://montreal-forced-aligner.readthedocs.io/en/latest/}}; otherwise, they are estimated using the Silero voice activity detection tool\footnote{\url{https://github.com/snakers4/silero-vad}}.  
Speaking duration is then defined as the sum of word durations and short pauses, excluding pauses longer than 0.6~s, which are treated as silence.
Speaking rate is computed as the total number of syllables divided by the speaking duration. Syllable counts are estimated based on vowel sequences: in Chinese, each character is treated as one syllable, whereas in English, French, German, and Spanish, syllables are approximated by counting contiguous vowel groups within words.
The RMS energy is computed from reverberant speech, averaged over active speech segments to account for speaking duration.
Since the mixtures include room acoustic simulations, the speaker-to-microphone distance is randomly sampled in a virtual room, with horizontal distances of 0.3–1.5 m to the microphone.
Language, gender, age, emotional state, and transcription are obtained directly from the original dataset annotations.

(2) Audio Mixtures:
Each corpus was partitioned into non-overlapping training, validation, and test subsets, ensuring disjoint speakers and no shared utterances across subsets. We generated 100,000 mixtures for training and 10,000 mixtures each for validation and testing. { The generated mixtures have variable durations of up to 6~s, resulting in 115.16~h of training audio, 10.97~h of validation audio, and 11.03~h of test audio.} To create realistic speech mixtures, each mixture is composed of two speech signals, $S_1$ and $S_2$, from different speakers. Both signals are first processed to remove initial and trailing silence, using forced alignment for transcribed speech or voice activity detection for untranscribed segments, and are subsequently temporally arranged with partial overlaps to mimic conversational dynamics. If either $S_1$ or $S_2$ is shorter than 3~s, the overlap duration is set to the length of the shorter signal, and zero-padding is applied at the beginning of the shorter signal using a randomly selected start offset uniformly drawn from $[0, \text{duration of longer sample} - \text{duration of shorter sample}]$. For longer signals, the overlap duration is set to $(\text{length}(S_1) + \text{length}(S_2) - 6)$~s, with $S_2$ padded using a fixed start offset of $6 - \text{length}(S_2)$~s.

To simulate realistic room acoustics, each $S_1$ and $S_2$ is convolved with a separate room impulse response (RIR) from the gpuRIR library\cite{diaz2021gpurir}, representing different speaker positions. Room dimensions are randomly sampled within 9--11~m (length/width) and 2.6--3.5~m (height), with RT60 values of 0.3--0.6~s. The microphone is positioned at the room center, and speakers are placed such that horizontal speaker–microphone distances range from 0.3 to 1.5~m and heights from 1.6 to 1.9~m. The reverberant $S_2$ is fixed while $S_1$ is scaled to a signal-to-interference ratio (SIR) uniformly sampled between –6 and 6~dB. For each mixture, one of the reverberant signals is randomly selected as the target speech.

{ (3) WHAM!-Noise Data Construction:}
{ To evaluate robustness to additive background noise while preserving the original relative-cue annotations, we construct noisy training, validation, and test splits using real ambient noise from the WHAM! dataset~\cite{wichern2019wham}. For each noise-free simulated two-speaker mixture, a WHAM! noise recording is selected from the corresponding split, downsampled to 16~kHz, and randomly cropped to the mixture duration. Training, validation, and held-out test noise recordings are used for the respective noisy splits. The noise segment is scaled to an SNR uniformly sampled from $[-6,10)$~dB, defined relative to the original two-speaker mixture, and then added to it. The noisy mixture is used as input, while the original target and interfering signals are retained for target-selection labels and signal-level evaluation. This procedure introduces real ambient noise without altering the original relative-cue labels or references.}

(4) Text Prompts: The details of text prompts generation are described in Section II. { In our controlled evaluation, prompts are generated from clean-speech annotations and measured attributes before mixing. This design provides unambiguous ground-truth cue labels and allows us to isolate the effect of cue representation.}
Table~\ref{tab:relative_cue_counts} summarizes the distribution of samples across relative cue categories in the training, validation, and test sets, excluding the ``Similar'' and ``Same'' category, as they do not offer discriminative power. Since the ``random'' and ``all'' cues are not available in every mixture to specify the target speaker, the number of samples associated with these cues does not correspond to the total number of mixtures.
\begin{table}[h]
    \centering
    \caption{Sample counts of each relative cue in the training, validation, and test sets, excluding cues labeled as ``Similar'' or ``Same''}
    \label{tab:relative_cue_counts}
    \renewcommand{\arraystretch}{1.2}
    \begin{tabular}{
    p{2.15cm}|
    >{\raggedleft\arraybackslash}p{1.2cm}|
    >{\raggedleft\arraybackslash}p{1.2cm}|
    >{\raggedleft\arraybackslash}p{1.2cm}
    }
        \hline
        \textbf{Relative Cue} & \textbf{Training} & \textbf{Validation} & \textbf{Test} \\
        \hline
        Random Cues 
        & 99641 & 9962 & 9956 \\ 
        \hline
        All Cues 
        & 99960 & 9996 & 9997 \\ 
        \hline
        Language
        & 73845 & 7192 & 7157 \\
        \hline
        Transcription 
        & 92311 & 9186 & 9189 \\ 
        \hline
        Gender 
        & 49114 & 5759 & 5125 \\ 
        \hline
        Emotion
        & 20377 & 1935 & 2012 \\ 
        \hline
        Speaking Rate
        & 67821 & 6674 & 6796 \\ 
        \hline
        Speaking Duration 
        & 76498 & 7880 & 7773 \\ 
        \hline
        Pitch 
        & 89366 & 8917 & 8909 \\ 
        \hline
        Pitch Range 
        & 77019 & 7715 & 7612 \\ 
        \hline
        Distance 
        & 27983 & 2672 & 2876 \\ 
        \hline
        Age
        & 3130 & 348 & 395 \\ 
        \hline
        Loudness 
        & 50731 & 4971 & 5008 \\ 
        \hline
        Temporal Order 
        & 60625 & 6686 & 6489 \\ 
        \hline
    \end{tabular}
\end{table}

\subsection{Training Configurations}
For the speech encoder in the classification model of the proposed two-stage framework, all cue types span linguistic, paralinguistic, and acoustic information, making the encoder central to capturing these diverse representations. We adopt the encoder from the pretrained \texttt{wav2vec2-large-xlsr-53} model, a multilingual self-supervised speech representation model based on the wav2vec 2.0 architecture~\cite{baevski2020wav2vec}. Trained on speech from 53 languages, including Multilingual LibriSpeech and other large-scale corpora, it learns universal, language-independent representations directly from raw audio. Compared to the ECAPA-TDNN speaker encoder, which focuses on speaker-specific embeddings, \texttt{wav2vec2-large-xlsr-53} captures richer linguistic, paralinguistic, and acoustic features, making it well suited for cross-lingual target speech extraction. We use the first five encoder layers, with fine-tuning applied to the fifth layer.

To encode text prompts into embeddings, we use the LLaMA 3.2 1B Instruct model for both single-stage TSE systems and the classification model in the two-stage framework. The LLaMA 3.2 1B instruction model is fine-tuned with LoRA (rank 16, scaling factor 16, dropout 0.05) on the query and key projection layers of self-attention.

To ensure high-quality signal separation, we adopt the TF-Locoformer~\cite{saijo2024tf}, which demonstrates state-of-the-art performance and robustness under reverberation. Single-stage TSE and two-stage separation models are trained with negative SI-SDR loss, float16 mixed precision, and a batch size of 24 (24 V100 GPUs, batch size 1 per GPU). The classification models also use float16 precision and batch size 24.

All models are optimized using AdamW \cite{LoshchilovH19}, with learning rates and gradient clipping configured individually for each model. The dual-path transformer (DPT)-based single-stage TSE model (used in the following subsection) was trained with a learning rate of { 3e-5}, a maximum gradient norm of 30, and 1000-step linear warm-up. The TF-Locoformer-based single-stage TSE was trained with a learning rate of { 3e-5} and a maximum norm of 5, whereas the two-stage separation model used a learning rate of 1e-3 with the same maximum norm. The classification model uses a learning rate of 1e-4 with max norm 30. Different learning rates and maximum norm values were adopted to account for the distinct architectures and convergence behaviors of the respective models. Learning rates are halved if validation loss does not improve for three epochs, and training employs early stopping with a maximum of 100 epochs.

\begin{table*}[t]
\centering
\caption{Comparison of single-stage baselines, text-guided two-stage selection, and oracle two-stage selection across different cues. Oracle selection chooses the separated output with the higher SI-SDR against the ground-truth target speech and is used only as an upper bound.}
\label{tab:single_two_stage_results}
\footnotesize
\setlength{\tabcolsep}{2.2pt}
\renewcommand{\arraystretch}{1.04}
\begin{tabular}{
>{\raggedright\arraybackslash}p{2.2cm} |
S[table-format=-1.1]
c |
S[table-format=-2.1]
c |
ccc |
c
c
}
\hline
\multirow{3}{*}{Cue Type} &
\multicolumn{4}{c|}{Single-Stage} &
\multicolumn{5}{c}{Two-Stage TF-Locoformer} \\
\cline{2-10}
& \multicolumn{2}{c|}{DPT-based TSE}
& \multicolumn{2}{c|}{TF-Locoformer TSE}
& \multicolumn{3}{c|}{Text-Guided Selection}
& \multicolumn{2}{c}{Oracle Selection} \\
\cline{2-10}
& \multicolumn{1}{c}{SI-SDRi} & \multicolumn{1}{c|}{PESQ}
& \multicolumn{1}{c}{SI-SDRi} & \multicolumn{1}{c|}{PESQ}
& {Sep ACC(\%)} & {SI-SDRi} & {PESQ}
& \multicolumn{1}{c}{SI-SDRi}
& \multicolumn{1}{c}{PESQ} \\
\hline
Random Cues       & 7.3  & 1.67 & 13.0 & 2.87 & 99.2 & 17.1 & 3.39 & 17.5 & 3.42 \\
All Cues          & 7.9  & 1.69 & 13.9 & 2.93 & 99.8 & 17.4 & 3.41 & 17.5 & 3.41 \\
Language          & 4.2  & 1.61 & 10.8 & 2.75 & 99.3 & 17.3 & 3.41 & 17.7 & 3.42 \\
Transcription     & 3.3  & 1.57 & 8.1  & 2.63 & 90.4 & 12.5 & 3.22 & 17.6 & 3.44 \\
Gender            & 7.6  & 1.68 & 12.0 & 2.81 & 98.1 & 16.7 & 3.37 & 17.7 & 3.42 \\
Emotion           & 3.9  & 1.49 & 7.2  & 2.53 & 87.2 & 10.3 & 3.10 & 17.2 & 3.41 \\
Pitch             & 6.6  & 1.65 & 11.1 & 2.77 & 94.6 & 14.8 & 3.30 & 17.6 & 3.42 \\
Pitch Range       & 4.2  & 1.58 & 7.4  & 2.56 & 88.7 & 11.3 & 3.14 & 17.6 & 3.41 \\
Loudness          & 7.6  & 1.69 & 13.6 & 2.90 & 97.2 & 16.0 & 3.35 & 17.5 & 3.41 \\
Distance          & 6.3  & 1.65 & 12.6 & 2.84 & 98.4 & 17.0 & 3.38 & 17.9 & 3.42 \\
Age               & -0.1 & 1.35 & -5.9 & 1.72 & 64.3 & -1.6 & 2.45 & 17.9 & 3.14 \\
Temporal Order    & 7.7  & 1.75 & 13.2 & 2.92 & 99.8 & 18.0 & 3.45 & 18.1 & 3.46 \\
Speaking Rate     & 3.1  & 1.59 & 6.1  & 2.54 & 91.3 & 13.1 & 3.27 & 17.7 & 3.46 \\
Speaking Duration & 6.4  & 1.69 & 10.7 & 2.76 & 95.9 & 15.6 & 3.34 & 17.8 & 3.44 \\
\hline
\end{tabular}
\end{table*}

\subsection{Two-Stage Method versus Single-Stage Baselines}
\textbf{Experimental Setup:}
We first evaluate the proposed two-stage method against single-stage baselines. In the first stage, the separation model is trained from scratch to obtain separated speech signals. In the subsequent classification stage, a single model is trained and evaluated using template prompts for each mixture, with each prompt corresponding to one of the fourteen relative cue types listed in Table~\ref{tab:relative_cue_counts}.  
The single-stage baselines include a DPT-based TSE model from our prior work \cite{dai25b_interspeech} and a TF-Locoformer-based TSE model, obtained by adapting the TF-Locoformer speech separation model for target speech extraction. Specifically, the model is reconfigured to predict only the target speaker by reducing the number of output speakers in the transposed convolution layer from 2 to 1. Text prompts categorized as ``same'' or ``similar'' are excluded from training, validation, and testing, as they provide limited discriminative power.

{ Recent single-stage text-guided TSE systems use different separation backbones, text encoders or fusion mechanisms between text embeddings and mixture embeddings. Therefore, their reported results are not directly comparable with our single-stage baselines, because differences in performance may come from any of these architectural choices rather than from the proposed two-stage paradigm itself. The purpose of the single-stage baselines in this paper is to provide a fair and controlled comparison. In particular, the TF-Locoformer-based single-stage TSE system uses the same separation backbone and text encoder as the proposed two-stage framework, so the comparison primarily reflects the mechanism difference between direct text-conditioned extraction and two-stage separation followed by text-guided selection.}

{ We include two enrollment-audio comparisons to position the proposed textual cues. The first is a controlled cue-modality comparison within the same two-stage pipeline: the separation front-end and separated candidates are fixed, while the selector uses either textual cues or enrollment audio. In the wav2vec2-large-xlsr-53 (hereafter referred to as wav2vec2)-based enrollment baseline, the text encoder is replaced by the same audio encoder used for the candidate signals. For each mixture, a clean utterance from the same target speaker is randomly selected, normalized to the mixture length by truncation or zero-padding, and convolved with a randomly drawn RIR to produce a reverberant enrollment utterance. Enrollment utterances are distinct from the corresponding mixtures, and evaluation uses mixture sets of the same size as those used for each textual cue.}

{ The second comparison uses a stronger enrollment-based selector with ECAPA-TDNN, a speaker-verification encoder optimized for speaker-identity embeddings. In this setting, ECAPA-TDNN extracts embeddings from both the enrollment utterance and the separated candidate signals, and target selection is performed by choosing the candidate with the higher cosine similarity to the enrollment embedding. This comparison provides a stronger enrollment-audio reference than the wav2vec2-based controlled baseline, while still sharing the same separated candidates rather than constituting a full benchmark against independently optimized enrollment-based TSE systems with different separators, architectures, or enrollment protocols.}

{ For the WHAM!-noise condition, we train a separate TF-Locoformer separation model on the noisy training mixtures and use the noisy validation split for model selection. Based on the separated noisy candidates, we train separate second-stage classifiers for textual-cue-based target selection and ECAPA-TDNN-based enrollment-audio target selection. Both second-stage classifiers are trained and evaluated under the same noisy data condition, enabling a direct robustness comparison between textual cues and the strong enrollment-audio selector.}

\textbf{Results and Analysis:}
We evaluate target speech extraction using scale-invariant signal-to-distortion ratio improvement (SI-SDRi) \cite{le2019sdr} and perceptual evaluation of speech quality (PESQ) \cite{941023}. For the TF-Locoformer-based two-stage framework, we report classification accuracy on separated speech (Sep ACC), text-guided TSE performance (SI-SDRi/PESQ), and an oracle-selection upper bound computed from the separated outputs. Results are summarized in Table~\ref{tab:single_two_stage_results}.

We evaluate three prompt types introduced in Section II (\textit{Individual}, \textit{All}, and \textit{Random}) on their corresponding mixtures. The two-stage framework also supports target speaker enrollment, selecting the separated signal with the highest cosine similarity to the enrollment embedding, enabling direct comparison with text-based cues. For all non-age cue types, the TF-Locoformer-based single-stage TSE system outperforms its DPT-based counterpart, achieving improvements of more than 3 dB in SI-SDRi and, on average, more than 1.0 in PESQ. The age cue remains challenging: TF-Locoformer improves PESQ but not SI-SDRi relative to DPT, indicating that age-based target specification is highly sensitive to cue discriminability and target-selection errors.
Classification performance across cue types reflects their contribution to single-stage baselines. Compared to single-stage TF-Locoformer-based TSE, the two-stage method shows clear gains across almost all cue types: the SI-SDRi improvement exceeds 3 dB for most cue types and remains substantial at 2.4 dB for loudness, while the average PESQ improvement exceeds 0.5. Even for age, the two-stage method improves over the TF-Locoformer single-stage result, although its absolute performance remains much lower than that of more discriminative cues. These results demonstrate the practical advantage of separating speech separation from text-guided target selection.

{ To isolate the contribution of text-guided target selection from the quality of the separator, Table~\ref{tab:single_two_stage_results} also reports an oracle-selection upper bound. This oracle uses the ground-truth target speech only for analysis and always selects the separated output with the higher SI-SDR with respect to the target. The gap between oracle and text-guided selection therefore quantifies the loss caused by non-oracle target-selection errors under the same separator. For all cues, random cues, language, and temporal order, the gap is small (0.1--0.4 dB SI-SDRi and at most 0.03 PESQ), because these cues are inherently more discriminative and therefore support more reliable target selection, allowing the text-guided classifier to nearly reach the separator upper bound. Larger gaps for age, emotion, pitch range, transcription, and speaking rate reflect weaker cue discriminability and lower classification accuracy rather than a limitation of the separation module itself.}

{ We next assess the reliability of classification decisions in the two-stage framework through confidence and calibration analysis. For each binary decision, confidence is defined as \(c=\max(\hat{y},1-\hat{y})\). As shown in Table~\ref{tab:confidence_calibration}, across 89294 test decisions, the classifier achieves an accuracy of 95.58\%, with correct decisions exhibiting substantially higher average confidence than incorrect ones (0.9150 vs.\ 0.6619). Empirical accuracy also increases consistently with confidence, and the overall expected calibration error is 0.0520, indicating that confidence is informative under the evaluated conditions. However, reliability remains cue-dependent: highly discriminative cues generally show high accuracy and low high-confidence error rates, whereas weaker cues exhibit poorer accuracy or calibration. Speaking duration is a notable exception, as some incorrect decisions remain highly confident despite its high overall accuracy. Overall, these results demonstrate that confidence provides a generally reliable indicator of prediction correctness under the evaluated conditions, although occasional high-confidence errors may still occur, particularly under unseen acoustic or speaker conditions.}

\begin{table}[t]
\centering

\caption{Cue-dependent confidence and calibration statistics for the second-stage classifier. HC wrong denotes the percentage of incorrect decisions with \(c>0.9\).}
\label{tab:confidence_calibration}
\scriptsize
\setlength{\tabcolsep}{1.6pt}
\renewcommand{\arraystretch}{1.02}
\begin{tabular}{>{\raggedright\arraybackslash}p{1.9cm}ccccc}
\toprule
Cue Type & \shortstack{ACC(\%)} & \(C_{\mathrm{corr}}\) &
\(C_{\mathrm{wrong}}\) & \shortstack{HC Wrong(\%)} & ECE \\
\midrule
Overall & 95.6 & 0.915 & 0.662 & 5.5 & 0.052 \\
Random Cues & 99.2 & 0.961 & 0.648 & 3.9 & 0.033 \\
All Cues & 99.8 & 0.984 & 0.614 & 0.0 & 0.015 \\
Language & 99.3 & 0.920 & 0.600 & 0.0 & 0.075 \\
Transcription & 90.5 & 0.891 & 0.674 & 5.2 & 0.035 \\
Gender & 98.1 & 0.932 & 0.639 & 2.1 & 0.055 \\
Emotion & 87.2 & 0.862 & 0.691 & 8.9 & 0.034 \\
Pitch & 94.6 & 0.864 & 0.622 & 4.2 & 0.095 \\
Pitch Range & 88.7 & 0.839 & 0.658 & 4.7 & 0.068 \\
Loudness & 97.2 & 0.921 & 0.646 & 2.9 & 0.059 \\
Distance & 98.4 & 0.905 & 0.619 & 0.0 & 0.084 \\
Age & 64.3 & 0.747 & 0.686 & 5.0 & 0.103 \\
Temporal Order & 99.9 & 0.960 & 0.670 & 0.0 & 0.039 \\
Speaking Rate & 91.4 & 0.863 & 0.657 & 2.2 & 0.068 \\
Speaking Duration & 96.0 & 0.920 & 0.708 & 19.1 & 0.048 \\
\bottomrule
\end{tabular}
\end{table}

\begin{figure}[htbp]
    \centering
    
    \includegraphics[width=0.39\textwidth]{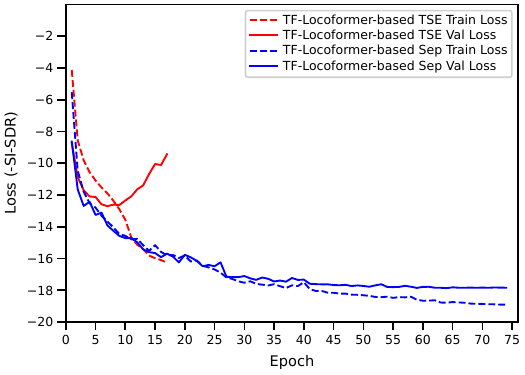}
    \caption{Training and Validation Loss Comparison for TF-Locoformer-based single-stage TSE and TF-Locoformer-based Separation models over epochs. The loss is measured such that lower (more negative) is better.}
    \label{fig:loss_comparison}
\end{figure}

We also present the training and validation loss curves of the TF-Locoformer-based single-stage TSE model and the TF-Locoformer-based separation model in Fig. 3. { From these curves, we observe that the single-stage TSE model is more prone to overfitting: its training loss continues to decrease while its validation loss deteriorates. In contrast, the separation model converges more gradually and achieves a lower validation loss. In the early training epochs, the separation model also obtains a lower training loss despite using a smaller learning rate. These observations suggest that the two-stage formulation provides a more stable optimization behavior.}
{ To examine whether this trend is representative rather than a single-run artifact, we conducted two additional independent training trials with different random seeds from the original run. A trial refers to a complete independent training run initialized with a different random seed. Both the TF-Locoformer-based single-stage TSE model and the TF-Locoformer-based separation model were trained under the same maximum epoch budget and the same early-stopping criterion, where training was stopped if the validation loss did not improve for 10 consecutive epochs. The same trend was observed in both additional trials: the TF-Locoformer-based single-stage TSE model consistently exhibited stronger overfitting than the TF-Locoformer-based separation model. This further supports the advantage of the proposed two-stage framework.}

{ Table~\ref{tab:classification_cue_results} summarizes the two enrollment-audio comparisons. For the original condition, the textual-cue columns report clean- and separated-speech classification accuracy, while the enrollment-audio columns include both the controlled \texttt{wav2vec2}-based baseline and the stronger ECAPA-TDNN-based baseline. For the WHAM!-noise condition, the table reports target-selection accuracy for textual cues and ECAPA-TDNN enrollment audio under the same noisy two-stage framework.}
\begin{table}[t]
\centering

\caption{Comparison of textual cues and enrollment-audio target selection under original and WHAM!-noise conditions. All entries are target classification/selection accuracy (ACC). w2v-C and w2v-S denote clean- and separated-speech wav2vec2-based enrollment baselines.}
\label{tab:classification_cue_results}
\scriptsize
\setlength{\tabcolsep}{1.2pt}
\renewcommand{\arraystretch}{1.02}
\begin{tabular}{@{}lccccccc@{}}
\toprule
\multirow{2}{*}{Cue} & \multicolumn{5}{c}{Original ACC(\%)} & \multicolumn{2}{c}{WHAM! ACC(\%)} \\
\cmidrule(lr){2-6}\cmidrule(l){7-8}
& Clean & Sep & w2v-C & w2v-S & \shortstack{ECAPA-\\TDNN} & Text. & \shortstack{ECAPA-\\TDNN} \\
\midrule
Random Cues & 99.3 & 99.2 & 95.7 & 95.7 & 98.8 & 98.9 & 98.1 \\
All Cues & 99.8 & 99.8 & 95.7 & 95.7 & 98.7 & 99.7 & 98.0 \\
Language & 99.5 & 99.3 & 96.6 & 96.6 & 99.5 & 98.7 & 99.0 \\
Transcription & 90.8 & 90.4 & 95.8 & 95.9 & 99.0 & 89.1 & 98.4 \\
Gender & 98.0 & 98.1 & 98.2 & 98.1 & 99.7 & 98.0 & 99.6 \\
Emotion & 85.8 & 87.2 & 88.4 & 88.6 & 98.3 & 85.8 & 96.6 \\
Pitch & 95.2 & 94.6 & 95.8 & 95.8 & 98.8 & 93.8 & 98.1 \\
Pitch Range & 88.4 & 88.7 & 95.9 & 96.0 & 98.8 & 87.1 & 98.0 \\
Loudness & 98.8 & 97.2 & 95.7 & 95.7 & 98.8 & 94.3 & 98.2 \\
Distance & 99.0 & 98.4 & 96.2 & 96.4 & 98.7 & 97.7 & 98.0 \\
Age & 64.0 & 64.3 & 93.4 & 93.4 & 94.2 & 57.7 & 92.4 \\
Temporal Order & 100 & 99.8 & 95.6 & 95.6 & 98.7 & 99.3 & 98.2 \\
Speaking Rate & 91.5 & 91.3 & 95.1 & 95.4 & 99.0 & 89.0 & 98.4 \\
Speaking Duration & 96.3 & 95.9 & 95.7 & 95.6 & 98.6 & 95.4 & 98.0 \\
\midrule
Average & 95.8 & 95.6 & 95.8 & 95.8 & 98.9 & 94.6 & 98.2 \\
\bottomrule
\end{tabular}
\end{table}

{ The original-condition results first show that clean- and separated-speech textual-cue accuracies are closely matched for most cue types, suggesting that the separated signals preserve the information needed for downstream target selection. Under the controlled \texttt{wav2vec2} enrollment setting, clean- and separated-speech enrollment results are also nearly identical, both averaging 95.8\%. This comparison isolates the conditioning modality under a shared two-stage pipeline and shows that textual cues can provide target-selection accuracy comparable to enrollment audio when the same general-purpose audio encoder is used.}

{ The stronger ECAPA-TDNN comparison gives a different perspective. When candidate and enrollment speech are encoded using ECAPA-TDNN, the average enrollment-audio accuracy increases to 98.9\% in the original condition. This is expected because ECAPA-TDNN is explicitly optimized for speaker verification, whereas \texttt{wav2vec2} captures a broader mixture of speaker identity, linguistic, prosodic, emotional, and acoustic-environment information. Therefore, the ECAPA-TDNN result can be regarded as a stronger enrollment-based two-stage reference. Even against this stronger baseline, several textual-cue configurations remain highly competitive: the random cues, all cues, and temporal order configurations achieve higher classification accuracy than ECAPA-TDNN-based enrollment audio, while the language and distance cues perform comparably. Other textual cues, including gender, loudness, speaking duration, pitch, speaking rate, and transcription, also achieve accuracies above 90\% under the original separated-speech condition.}

{ Under the WHAM!-noise condition, both textual cues and ECAPA-TDNN-based enrollment audio remain robust. The average textual-cue accuracy decreases only from 95.6\% to 94.6\%, while the ECAPA-TDNN enrollment accuracy decreases from 98.9\% to 98.2\%. The random cues, all cues, and gender configurations exhibit particularly strong noise robustness, while language, distance, and temporal order also remain robust. In addition, gender, loudness, speaking duration, and pitch continue to achieve accuracies above 90\%. These results demonstrate that textual cues provide a practical and valuable alternative, particularly when enrollment speech is unavailable or avoided because of privacy concerns.}

\begin{figure}[htbp]
    \centering
    \includegraphics[
        width=0.97\columnwidth,
        height=0.68\textheight,
        keepaspectratio
    ]{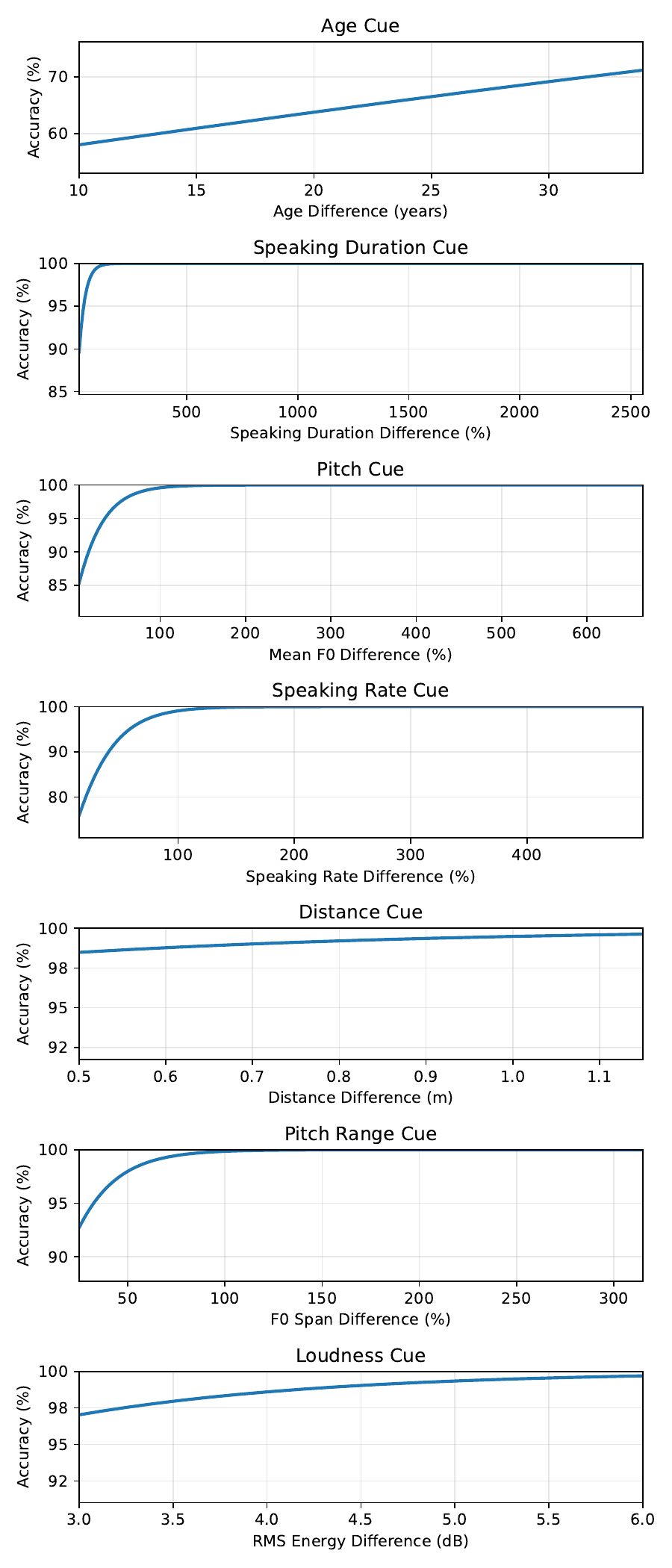}
    \caption{Classification accuracy as a function of the attribute difference between target and interfering speakers for each cue type.}
\end{figure}

We further analyze classification accuracy as a function of the continuous-valued attribute difference between paired target–interference clean speech samples for each textual cue type. The temporal order cue is excluded due to its perfect classification accuracy (100\%). For the remaining cues, logistic regression is used to obtain a smooth estimate of overall classification accuracy over the continuous attribute space. As shown in Fig. 4, accuracy exhibits a clear increasing trend as attribute differences grow. Except for age, all cues reach near-perfect classification when the differences become sufficiently large.

\subsection{Relative Cues versus Independent Cues}
\textbf{Experimental Setup:}
We then evaluate whether relative cues outperform independent cues for target speech extraction within a two-stage framework. For discrete-valued attributes, relative and independent cues provide the same level of discriminability; therefore, we restrict our evaluation to originally continuous-valued attributes. In total, eight cues are considered, of which two---\textit{age} and \textit{temporal order}---are excluded.
The \textit{age} cue is excluded because the training data are limited to speakers aged 18--50, making absolute age categorization unreliable. Applying common-sense age groupings (young: $<25$, youth: 25--40, middle: 41--64, senior: $>65$) would lead to severe class imbalance (young: 7063; youth: 10310; middle: 2627; senior: 0). The \textit{temporal order} cue, defined as the difference in onset time between the target and interfering speech, is inherently relational and therefore does not admit a meaningful absolute categorization.
Consequently, the evaluation focuses on the remaining six cues: \textit{pitch}, \textit{pitch range}, \textit{loudness}, \textit{distance}, \textit{speaking rate}, and \textit{speaking duration}.

Separated speech samples are selected from the TF-Locoformer separation model based on mixtures associated with these cue types. Two binary classification models are trained: one guided by relative-cue text prompts and the other by independent-cue text prompts.

\paragraph{Independent Cues} Independent cue categories are defined based on the distribution of attribute values in the training samples (including both target and interference speech), following previous studies\cite{huo2025beyond, jiang2025listen}. We use equal-frequency binning, where each attribute is quantized into two or three categories with similar amount of samples per each category:

\begin{itemize}
    \item Pitch: low / normal / high
    \item Pitch range: narrow / normal / wide
    \item Loudness: quiet / normal / loud
    \item Distance: near / far
    \item Speaking rate: slow / normal / fast
    \item Speaking duration: short / long
\end{itemize}

\paragraph{Relative Cues} Relative cue categories are defined based on the process explained in Section~II and the thresholds specified in Table~I:

\begin{itemize}
    \item Pitch: lower / similar / higher 
    \item Pitch range: narrower / similar / wider
    \item Loudness: quieter / similar / louder 
    \item Distance: nearer / similar / farther 
    \item Speaking rate: slower / similar / faster
    \item Speaking duration: shorter / similar / longer
\end{itemize}
\begin{table*}[htbp]
\centering
\caption{Metrics for Independent and Relative Cues}
\label{tab:independent_relative_metrics}
\setlength{\tabcolsep}{4pt}
\renewcommand{\arraystretch}{0.94}
\begin{tabular}{
  ll
  r r S[table-format=-2.1] S[table-format=1.2]
  r r S[table-format=-2.1] S[table-format=1.2]
}
\toprule
& &
\multicolumn{4}{c}{\textbf{Independent Cues}} &
\multicolumn{4}{c}{\textbf{Relative Cues}} \\
\cmidrule(lr){3-6} \cmidrule(lr){7-10}
Cue Type & Group
& \multicolumn{1}{c}{N}
& \multicolumn{1}{c}{ACC(\%)}
& \multicolumn{1}{c}{SI-SDRi}
& \multicolumn{1}{c}{PESQ}
& \multicolumn{1}{c}{N}
& \multicolumn{1}{c}{ACC(\%)}
& \multicolumn{1}{c}{SI-SDRi}
& \multicolumn{1}{c}{PESQ} \\
\midrule

Overall & non-same / non-similar
& \Nperc{34501}{58.1} & 93.5 & 14.3 & 3.29
& \Nperc{\underline{38974}}{66.9} & 94.1 & 14.5 & 3.29 \\

Overall & all
& \Nperc{59009}{100} & 75.3 & 4.3 & 2.87
& \Nperc{59009}{100} & 79.2 & 6.3 & 2.93 \\
\midrule
Distance & non-same / non-similar
& \Nperc{5230}{52.3} & 94.0 & 14.4 & 3.29
& \Nperc{2876}{28.8} & 98.5 & 17.0 & 3.38 \\

Distance & all
& \Nperc{10000}{100} & 72.9 & 2.9 & 2.81
& \Nperc{10000}{100} & 64.1 & -2.1 & 2.53 \\

\midrule
Loudness & non-same / non-similar
& \Nperc{6994}{69.9} & 91.8 & 13.0 & 3.23
& \Nperc{5008}{50.0} & 97.9 & 16.3 & 3.36 \\

Loudness & all
& \Nperc{10000}{100} & 79.2 & 6.1 & 2.94
& \Nperc{10000}{100} & 74.4 & 3.7 & 2.85 \\

\midrule
Pitch & non-same / non-similar
& \Nperc{6630}{66.3} & 98.1 & 16.8 & 3.39
& \Nperc{8909}{94.7} & 95.4 & 15.2 & 3.32 \\

Pitch & all
& \Nperc{9990}{100} & 82.4 & 8.2 & 3.03
& \Nperc{9990}{100} & 90.4 & 12.4 & 3.18 \\

\midrule
Pitch Range & non-same / non-similar
& \Nperc{6652}{66.5} & 89.7 & 12.0 & 3.19
& \Nperc{7612}{76.2} & 89.6 & 11.9 & 3.17 \\

Pitch Range & all
& \Nperc{9988}{100} & 77.0 & 5.0 & 2.90
& \Nperc{9988}{100} & 80.0 & 6.5 & 2.93 \\

\midrule
Speaking Rate & non-same / non-similar
& \Nperc{5979}{64.4} & 90.8 & 13.0 & 3.26
& \Nperc{6796}{75.3} & 89.8 & 12.3 & 3.23 \\

Speaking Rate & all
& \Nperc{9031}{100} & 75.9 & 4.9 & 2.91
& \Nperc{9031}{100} & 80.1 & 7.0 & 2.99 \\

\midrule
Speaking Duration & non-same / non-similar
& \Nperc{3016}{30.2} & 99.5 & 18.6 & 3.50
& \Nperc{7773}{77.7} & 96.7 & 16.0 & 3.36 \\

Speaking Duration & all
& \Nperc{10000}{100} & 64.2 & -1.2 & 2.61
& \Nperc{10000}{100} & 86.3 & 10.4 & 3.10 \\

\bottomrule
\end{tabular}
\end{table*}
We use audio mixtures from the dataset described in Section~IV that are annotated with the six relative cues—pitch, pitch range, loudness, distance, speaking rate, and speaking duration. 
We reuse the same predefined templates structure as described in Section II to generate prompts for independent cues. To isolate the effect of cue representation, we control for linguistic variability by regenerating relative-cue prompts using the same predefined templates as those used for independent cues. This ensures that both prompt types share identical sentence structure and phrasing, differing only in whether the cue is expressed in relative or independent terms. Consequently, any performance differences can be attributed specifically to the cue formulation rather than to variations in prompt wording or style. For example, a regenerated relative-cue prompt for a given target speech sample is ``Please extract the speaker with a higher pitch in the audio,'' whereas the corresponding independent-cue prompt is ``Please extract the speaker with a high pitch in the audio.''

During training, pairs of audio samples and prompts in which both speech signals belong to the same absolute category or the ``similar'' relative category are excluded. During evaluation, all pairs of audio samples and prompts are retained, with target labels derived from the PIT results of the separation model.
Classification accuracy is reported under different conditions. For independent cues, three conditions are considered: (i) same category, (ii) adjacent categories (e.g., slow–normal for speaking rate), and (iii) distinctly different categories (e.g., slow–fast for speaking rate). For relative cues, two conditions are considered: (i) same category (``similar'') and (ii) different categories. We further analyze cases in which target–interference pairs fall into different relative-cue categories but are quantized into the same, adjacent, or distinct absolute categories, highlighting scenarios where independent cues provide limited discriminability while relative cues remain informative.

\textbf{Results and Analysis:} We compare overall classification accuracy across cue types to evaluate the effectiveness of relative versus independent cues for TSE. For independent cues, pairs in the same category are labeled ``same,'' and those in different categories ``non-same.'' For relative cues, pairs in the same category are labeled ``similar,'' and those in different categories ``non-similar.'' Results are reported over six continuous-valued attributes.

For independent cues, we report classification accuracy along with SI-SDRi and PESQ for the ``non-same'' and ``all'' groups. For relative cues, we report these metrics for the ``non-similar'' and ``all'' groups. We explicitly account for the number of samples (N) in each group to fairly compare whether relative cues can exploit more information than independent cues. Notably, each relative cue and its corresponding independent cue are evaluated using the same number of samples. The results are summarized in Table~\ref{tab:independent_relative_metrics}.

The overall classification accuracy, SI-SDRi, and PESQ achieved with relative cues surpass those obtained with independent cues. Moreover, the number of samples in the non-similar group is larger than that in the similar group, highlighting the advantage of relative cues in leveraging more discriminative information. In particular, for pitch, pitch range, speaking rate, and speaking duration, the relative categories demonstrate a clear advantage over absolute categories in utilizing more distinguishing information.
\begin{figure}[!t]
    \centering
    \subfloat[{\textrm{\fontsize{9}{10}\selectfont Independent Speaking Rate Cue}}]{%
        \includegraphics[width=0.67\columnwidth]{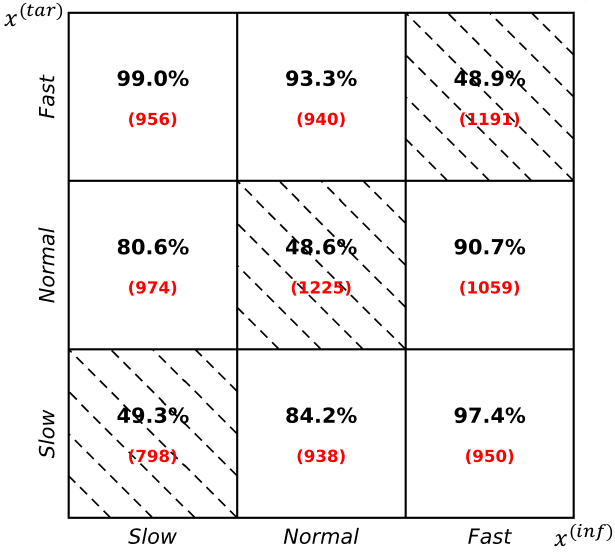}%
    }\\
    \subfloat[{\textrm{\fontsize{9}{10}\selectfont Relative Speaking Rate Cue}}]{%
        \includegraphics[width=0.67\columnwidth]{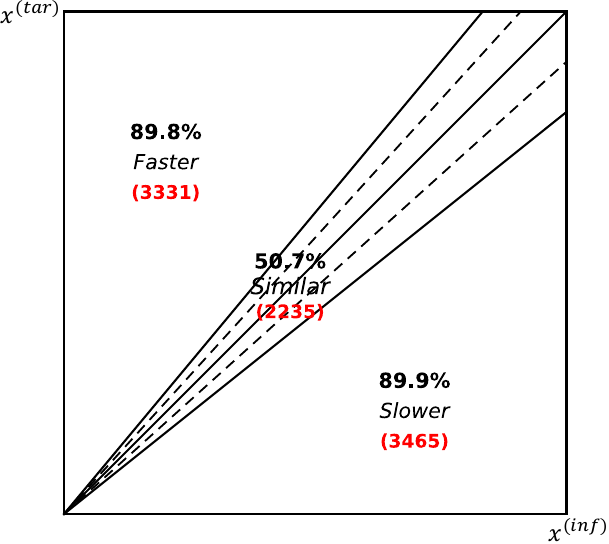}%
    }
    \caption{Classification accuracy and corresponding sample distribution across regions for independent (a) and relative (b) speaking rate cues, where $x^{(\mathit{tar})}$ (y-axis) and $x^{(\mathit{inf})}$ (x-axis) represent the speaking rate values of the target and interfering speakers, respectively.} 
    \label{fig:speaking_rate_regions}
\end{figure}
For the distance and loudness cues, using relative categories assigns more samples to the ``similar'' group than absolute categories. This is due to the limited ranges of simulated speaker-to-microphone distances and SIR values. In our dataset, distances range from 0.3 to 1.5~m, with a 0.5~m threshold for distinguishing speakers, leading many pairs to be classified as having similar distances. A similar effect is observed for the loudness cue: loudness is approximated using SIR, which is constrained to the range of -6 to 6 dB in our simulation. With a loudness cue threshold of 3 dB, nearly half of the samples are categorized as having similar loudness.

{ To quantify this effect, we provide a threshold-sensitivity analysis for loudness and distance. The tested thresholds were chosen for three purposes: the original perceptually conservative threshold (3 dB for loudness and 0.5 m for distance), a smaller threshold that makes the three relative categories approximately balanced (2 dB and 0.2 m), and an even smaller threshold that makes the ``similar'' category clearly the smallest (1 dB and 0.1 m). For each cue–threshold setting, we trained an independent text-guided classifier and evaluated both classification accuracy and TSE performance on the 10,000-sample full test set. The results are reported in Table~\ref{tab:threshold_sensitivity}.}

\begin{table}[t]
\centering

\caption{Threshold sensitivity of loudness and distance relative cues on the 10,000-sample full test set. NS denotes the combined non-similar categories (louder/quieter for loudness and nearer/farther for distance). Bold thresholds indicate the original values used in the main experiments.}
\label{tab:threshold_sensitivity}
\scriptsize
\setlength{\tabcolsep}{2.0pt}
\renewcommand{\arraystretch}{1.05}
\begin{tabular}{@{}llccccccc@{}}
\toprule
Cue & Thr. & NS & Sim. & \multicolumn{3}{c}{ACC(\%)} & SI-SDRi & PESQ \\
\cmidrule(lr){5-7}
 & & Count & Count & NS & Sim. & All & & \\
\midrule
Loudness & 1 dB & 8364 & 1636 & 94.4 & 50.9 & 87.3 & 10.6 & 3.13 \\
Loudness & 2 dB & 6697 & 3303 & 96.8 & 50.3 & 81.5 & 7.4 & 3.00 \\
Loudness & \textbf{3 dB} & 5002 & 4998 & 98.2 & 50.0 & 74.1 & 3.6 & 2.85 \\
\midrule
Distance & 0.1 m & 8322 & 1678 & 90.9 & 49.4 & 83.9 & 8.9 & 3.06 \\
Distance & 0.2 m & 6599 & 3401 & 94.9 & 49.6 & 79.5 & 6.5 & 2.95 \\
Distance & \textbf{0.5 m} & 2876 & 7124 & 98.7 & 49.7 & 63.8 & -2.1 & 2.57 \\
\bottomrule
\end{tabular}
\end{table}

{ Table~\ref{tab:threshold_sensitivity} shows a clear trade-off between cue discriminability and coverage. Increasing the threshold improves classification of the non-similar categories because the retained `louder/quieter'' or `nearer/farther'' samples exhibit more salient differences. However, it also assigns a larger proportion of samples to the non-discriminative ``similar'' category, thereby reducing overall TSE performance on the full test set. This effect is especially pronounced for distance under the original threshold, for which most samples are labeled as similar.}
{ The results support the conclusion that the original thresholds produce reliable non-similar cues but limit cue coverage. In practice, loudness and distance cues should be used only when the perceived differences are clearly discriminative; otherwise, users should be prompted to provide alternative relative cues.}

It is informative to examine the specific categories of target–interference samples. Fig.~5 presents the classification accuracy and sample distribution for both independent (Fig.~5(a)) and relative (Fig.~5(b)) cues, using speaking rate as an example. Compared with the relative speaking rate cue, the independent cue results in more samples being categorized into the ``same'' group (slow–slow, normal–normal, fast–fast), which cannot be distinguished. In contrast, the relative cue preserves the fine-grained distinctions among these samples, supporting the assumption in Section~II.
For the independent speaking rate cue, accuracy is highest when target and interference fall into clearly distinct categories (e.g., slow--fast or fast--slow), and decreases for adjacent categories (e.g., slow--normal or normal--fast), as smaller relative differences make classification more difficult.
\begin{figure}[htbp]
\centering
\includegraphics[width=0.42\textwidth,height=0.21\textheight,keepaspectratio]{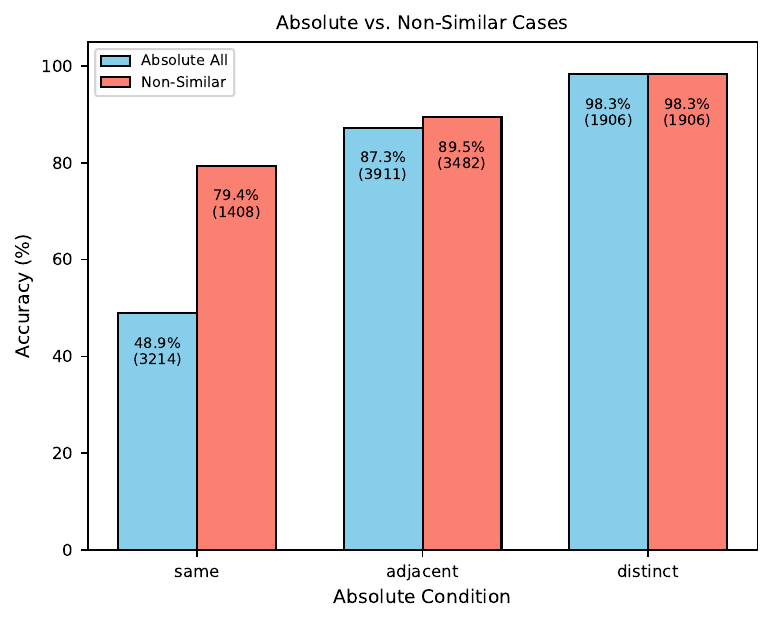}
\caption{Comparison between independent and relative cue categorizations for the speaking rate attribute.}
\end{figure}
\begin{figure}[!t]
    \centering
    \subfloat[{\textrm{\fontsize{9}{10}\selectfont Independent Speaking Rate Cue}}]{%
        \includegraphics[width=0.4\textwidth]{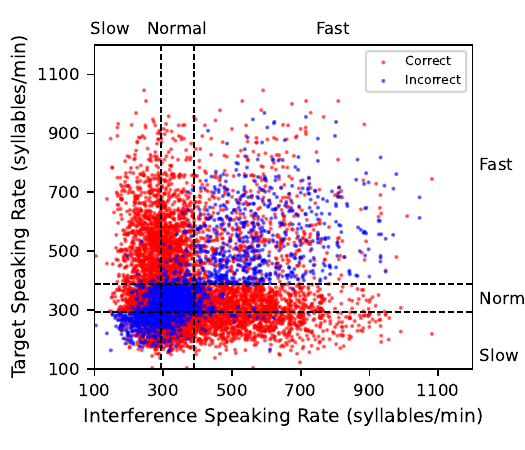}%
        \label{fig:cue_absolute}
    }
    \hfill
    \subfloat[{\textrm{\fontsize{9}{10}\selectfont Relative Speaking Rate Cue}}]{%
        \includegraphics[width=0.4\textwidth]{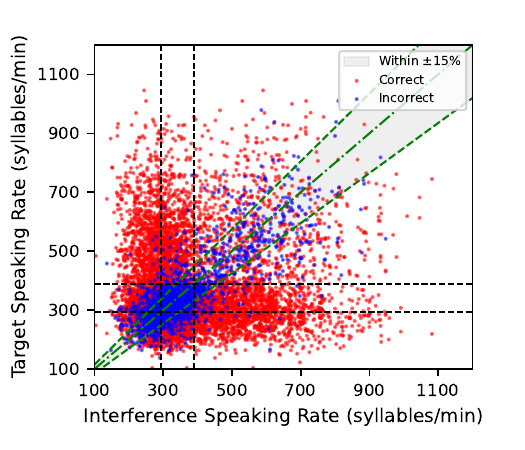}%
        \label{fig:cue_relative}
    }
    \caption{Joint distribution of classification accuracy for independent (a) and relative (b) speaking-rate cues across all paired target--interference samples. The x-axis and y-axis correspond to the speaking rates (syllables per minute) of the interfering and target speakers, respectively. Red points indicate correctly classified samples, while blue points indicate incorrectly classified samples. { The plotted speaking-rate set contains 9031 paired samples: independent cues have 6855 correct and 2176 incorrect samples (points), whereas relative cues have 7234 correct and 1797 incorrect points. The gray region denotes target--interference speaking-rate differences within $\pm 15\%$ and contains 2235 points.}}
    \label{fig:speaking_rate_joint_distribution}
\end{figure}

To further analyze the relationship between independent and relative cue categorizations, we focus on target--interference pairs labeled as ``non-similar'' and examine whether they fall into the same, adjacent, or distinct absolute categories. Using speaking rate as an example (Fig.~6), we report classification accuracy for these conditions on both the full set and the relative-different subset, along with their sample distribution. Accuracy increases with speaking rate difference: in the same condition, it is 48.9\% but improves to 79.4\% for ``non-similar'' pairs, indicating added discriminative power from relative comparison. In the adjacent condition, accuracy rises from 87.3\% to 89.5\%, while in the distinct condition it reaches 98.3\% in both cases. These results show that performance strongly correlates with speaking rate difference, with relative cues particularly benefiting cases where absolute quantization obscures pairwise distinctions.

Fig. 7 further illustrates the joint distribution of target and interference speaking rates under absolute and relative categorizations. Compared with independent cues, relative cues yield more correct classifications in the slow--slow, normal--normal, and fast--fast regions, highlighting their advantage when absolute categories are identical.
{ \section{Discussion}
In our implementation, the proposed two-stage TSE system incurs a higher computational cost than the single-stage baseline in terms of parameter count and inference time. This increase is primarily attributable to the wav2vec2-based audio encoder. However, this observation is specific to the present implementation and should not be regarded as an inherent limitation of two-stage architectures. In principle, two-stage systems can be more computationally efficient than single-stage systems by confining cross-modal processing to a specific component while allowing the remaining modules to operate unimodally and potentially more efficiently. The current implementation is therefore intended for offline evaluation, prioritizing extraction accuracy and interpretability rather than real-time or energy-efficient deployment. Moreover, text-guided TSE is inherently less suited to strictly online or streaming scenarios, as perceptual prompts often require users to listen to the mixture and compare speakers before specifying the target. The proposed approach is thus better suited to offline, interactive, or repeated-listening applications. Achieving streaming operation would require further developments, such as causal separation, incremental source tracking, stable stream assignment, and interactive prompting that allows users to update the target cue after comparing the speakers.

In practical use, listeners may not know the exact clean-speech attributes in advance. However, relative cues do not require users to estimate physical values such as mean \(F_0\) or speaking rate; users can listen to the mixture one or more times and compare the speakers perceptually, selecting cues that are clearly discriminative, such as the speaker with higher pitch, lower loudness, or slower speaking rate. An interactive system could also guide the user to compare speakers in a comparative way before entering a prompt. If the provided prompt is inaccurate or mismatched to the actual mixture, the classifier may select the wrong separated candidate, especially when the attribute difference is small. Providing reliable cue suggestions through the TSE system, allowing repeated listening, and asking users to revise ambiguous cues are practical strategies for reducing such mismatches. Nevertheless, the problem of incorrect or ambiguous prompts is beyond the scope of this work and is left for future work.

Although this paper focuses on two-speaker mixtures, the target-selection mechanism could be extended to mixtures containing more than two speakers by ranking the separated candidates according to their text–speech similarity scores. In practical applications, prompts should preferably describe salient extremal attributes, such as the speaker with the highest or lowest pitch or the fastest or slowest speaking rate, rather than requiring users to rank all speakers. More detailed prompts, such as selecting the speaker with the second-highest pitch, are also possible but may be less practical when many speakers are present. Scenarios involving an unknown number of speakers would additionally require source-count estimation, separation or diarization, and ranking among multiple candidate sources. These extensions are beyond the scope of the present work and are left for future research.

For an audio sample containing only one speaker throughout, possibly with background noise, there is no target-versus-interferer speaker ambiguity to resolve. Such cases therefore fall outside the core TSE problem and are closer to speech enhancement or denoising. In a practical pipeline, they can be handled by adding a source-counting or overlap-detection front end: if only one speaker is detected, the system can directly output or enhance that source; if multiple speakers are detected, the proposed two-stage separation-and-selection framework is applied.}

\section{Conclusion}
We have presented a comprehensive study on text-guided target speech extraction that leverages inter-speaker relative cues across a wide range of speech attributes. This is achieved through theoretical analysis, redesigning the model architecture with a two-stage separation-and-classification approach, and empirically evaluating relative versus independent cues on comparable continuous-valued attributes.

{ Experimental results show that the proposed two-stage approach substantially improves SI-SDRi and PESQ over single-stage text-based TSE systems. Cue-level analyses indicate that, although cues vary in discriminative power, several textual-cue settings are competitive with enrollment-audio selection in controlled two-stage comparisons. Specifically, the all cues, random cues, and temporal order settings outperform both the wav2vec2-based and strong ECAPA-TDNN-based enrollment-audio selectors, while the language and distance cues remain highly competitive. The WHAM!-noise evaluation demonstrates the strong robustness of the textual cues to real ambient additive noise, with the all cues, random cues, and gender settings being particularly robust. These findings establish textual relative cues as a practical alternative especially when enrollment speech is unavailable or raises privacy concerns. Furthermore, our results confirm that relative cues generally outperform independent cues, underscoring their advantage for text-guided target speech extraction. Notably, adapting a state-of-the-art separation model into a single-stage TSE system can also significantly enhance target speech extraction, further validating the effectiveness of incorporating advanced separation models into TSE frameworks.}

Overall, the findings indicate that integrating perceptually motivated relative cues with a two-stage model provides an effective and flexible framework for text-guided TSE. The proposed two-stage framework not only mitigates the speaker confusion inherent
in single-stage systems but also provides valuable insights into the relative importance of different speech attributes for TSE. {In future work, we plan to address the limitations discussed in Section V by extending the framework to mixtures with unknown or more than two speakers, reliable relative cues recommendation, and by exploring interactive or streaming-compatible prompting strategies.}

\bibliographystyle{IEEEtran}
\bibliography{mybib}

\end{document}